\documentclass{article}
\usepackage{amsfonts,amssymb, amsmath}
\usepackage[english]{babel}

\def\bq{ \begin{equation}}
\def\eq{ \end{equation}}
\def\ben{ \begin{eqnarray}}
\def\en{ \end{eqnarray}}

\begin{document}

%%%%%%%%%%%% TITLE %%%%%%%%%%%%%%

\title{On integrable systems outside  Nijenhuis and Haantjes geometry}
\author{  A. V. Tsiganov,\\
\it\small
St.Petersburg State University, St.Petersburg, Russia\\
\it\small e-mail: andrey.tsiganov@gmail.com}

\date{}
\maketitle

\begin{abstract}
We study non-invariant Killing tensors with non-zero Nijenhuis torsion in the three-dimensional Euclidean space.
Generalizing the corresponding integrable systems we construct  two new families of superintegrable systems in $n$-dimensional Euclidean space.
 \end{abstract}


%%%%%%%%%%%%%%%%%%%%%%%%%%%%%%%%%%%%%%%%%
\section{Introduction}
\setcounter{equation}{0}
Let us consider  Riemannian or pseudo-Riemannian manifold  $M$, dim$M=n$ endowed with coordinates $q=(q_1,\ldots,q_n)$. Metric  $\mathrm g(q)$ and potential $V(q)$ define the Hamilton function on  the  cotangent bundle $T^*M$
\bq\label{ham-g}
H_1=\sum_{i,j=1}^n \mathrm g^{ij}(q) p_i p_j+V(q)\,.
\eq
A criterion for orthogonal separability of the corresponding Hamilton-Jacobi equation is given by Benenti  \cite{ben16}:\\
The  Hamiltonian system defined by (\ref{ham-g}) is orthogonally separable if and
only if there exists a valence-two symmetric Killing tensor $K$ with
\begin{enumerate}
  \item pointwise simple and real eigenvalues with respect to metric  $\mathrm g$;
  \item orthogonally integrable eigenvectors with respect to metric $\mathrm g$;
  \item  such that
  \bq\label{kdv}
  d(K dV ) = 0.\eq
\end{enumerate}
A Killing tensor satisfying conditions  (1-2) is called a characteristic Killing tensor.

Thus, we have a family of the St\"{a}ckel systems which are  completely defined by Hamiltonian $H_1$  (\ref{ham-g}) and yet another integral of motion
\bq\label{int-st}
H_2=\sum_{i,j=1}^n K^{ij} p_i p_j+U(q)
\eq
involving  characteristic Killing tensor. Other $n-2$ independent integrals of motion in the involution $H_2,\ldots,H_n$ are defined by recurrence relations \cite{ben16,eis34,kal86}, all these integrals are polynomials of second order in momenta. Similar construction of $n-2$ additional integrals in more generic case including non-potential forces was proposed by  Kozlov \cite{koz20}.

The integrability of eigenvectors condition (2) is equivalent to a system of non-linear partial differential equations which can be represented in various forms, see \cite{smir05} and references within. For instance, a given symmetric Killing tensor $K$ has integrable eigenvectors if
\bq\label{nts}
[N^\ell_{[jk}\mathrm g_{i]\ell} = 0,\quad N^\ell_{[jk}K_{i]\ell} = 0,\quad N^\ell_{[jk}K_{i]m}K^m_\ell=0\,,
\eq
 where the square brackets stand for antisymmetrisation and $N$ denotes the Nijenhuis torsion of (1,1)  tensor field $K$,
 i.e. $N=N_K$ \cite{nij51}  .

According to  Haantjes  \cite{haa55}, these partial differential equations may be rewritten in the equivalent form
\bq\label{nh-zero}
 H_K=0
  \eq
where $H_K$ is the Haantjes  torsion of (1,1)  tensor field $K$.  Now Nijenhuis and Haantjes tensors can be found in various parts of mathematics,  mathematical physics, and classical mechanics, but the overwhelming majority of applications is related to the vanishing of one of these tensors \cite{bog96,bol19,smir05,tt21}.

In \cite{eis34} Eisenhart found eleven characteristic Killing tensors $K$ in  the three-dimensional Euclidean space $\mathbb R^3$ so that
\[
N_K(u,v)\neq 0\,,\qquad \mbox{but}\qquad H_K(u,v)=0
\]
for any vector fields $u$ and $v$, and proposed construction of the completely integrable St\"ackel systems associated with this tensor $K$.

Invariant construction of these eleven Killing tensors in $\mathbb R^3$ is discussed in  \cite{smir05}. For other existing Killing tensors  $K$ in the three-dimensional Euclidean space $\mathbb R^3$ we have
 \bq\label{nh-nz}
 N_K\neq 0\qquad\mbox{and}\qquad H_K\neq 0\,.
 \eq
 If such tensor $K$ is invariant under action a Killing vector field $X$
 \[
 H_K\neq 0\,,\qquad\mbox{but}\qquad \mathcal L_X\,K=0\,,
 \]
 where $\mathcal L$ is a Lie derivative and $X$ belongs to the isometry group $\mathbb E(3)$, then we also have an integrable system with integrals of motion
  $H_{1,2}$ (\ref{ham-g}-\ref{int-st}) and
 \[
 H_3=\sum_{i=1}^n X^i p_i \,,
 \]
 Thus, thanks to Noether's theorem, we have an integrable Hamiltonian system outside   Nijenhuis and Haantjes geometry.

 In \cite{ts15,ts15a} we found two  completely non-invariant (1,1) Killing tensors $K$  in the three-dimensional Euclidean space $\mathbb R^3$ so that
 \[
 H_K\neq 0\qquad\mbox{and}\qquad \mathcal L_X\,K\neq0\,,\qquad \forall X\in\mathbb E^3\,.
 \]
 The corresponding integrable Hamiltonian systems admit two with integrals of motion $H_{1,2}$ (\ref{ham-g}-\ref{int-st}), which are polynomials of second order in momenta,
 and one integral of motion  which is a polynomial of fourth order in momenta
 \[
 H_3=\sum_{i,j,k,\ell=1}^3A^{ijk\ell} p_ip_jp_kp_\ell+\sum_{i,j=1}^3 S^{ij} p_i p_j+W(q) \,.
 \]
 In this note, we obtain a few generalizations of these Killing tensors in  $\mathbb R^4$ that allows us  to describe  two families of superintegrable  Hamiltonian systems
in $\mathbb R^n$ outside of the  Nijenhuis and Haantjes geometry. All these systems have $n-1$ quadratic and one quartic integrals of motion in the involution.

\subsection{Killing, Nijenhuis and Haantjes tensors}
There are a few equivalent definitions of the Killing tensors \cite{benn06}.

For instance,  a Killing tensor $K$ of valence $p$ defined in $(M, \mathrm g)$ is a symmetric $(m, 0)$
tensor satisfying the Killing tensor equation
\bq\label{k-tens}
 [\![K,\mathrm g ]\! ] = 0
 \eq
where $[\![,]\! ]$ denotes the Schouten bracket. When $m = 1$, vector field $K=X$ is said to be a Killing
vector (infinitesimal isometry) and this equation reads as
\bq\label{isom}
\mathcal L_X\,\mathrm g = 0
\eq
where $L$ denotes the Lie derivative operator.

According to another definition, if  $q_i(t)$ is a geodesic, then   $K$ is a $(m,0)$  Killing tensor if scalar
 \[C =\sum K^{i_1,\ldots,i_m}\, p_{i_1}\cdots p_{i_m}\]
 is constant along a geodesic. Here $p_i=\dot{q}_i(t)$ is the tangent vector of the geodesic.

At $m=2$ we do not separate the motion along geodesics and the motion in a potential field, so we consider Killing tensor $K$ as a solution of the  equation
 \[
 \{H_1,H_2\}=0
 \]
 where
 \[ H_1=\sum_{i,j=1} \mathrm g^{ij}(q) p_i p_j+V(q)\,\qquad\mbox{and}\qquad H_2=\sum_{i,j=1} K^{ij} p_i p_j+U(q)\,,\]
 with "non-trivial" potentials $V(q)$ and $U(q)$. Below we explain  what means "non-trivial" potentials.

Metric establishes an isomorphism between the tangent space and its dual. This identifies co- and contravariant tensor components via
lowering or rising indices using the metric. In particular a tensor $K$ of valency two can be identified with  (0, 2), (2,0), or (1, 1) tensor field.

If $K$ be  (1,1)  tensor field in $M$, then  Nijenhuis and  Haantjes tensors on  $M$ are
\[
N_K(u,v)= K^2[u,v]+[Ku,Kv]-K\left([Ku,v]+[u,Kv]\right)\,,
\]
and
\[
 H_K(u,v)=K^2N_K(u,v)+N_K(Ku,Kv)-K\left( N_K(Ku,v)+N_K(u,Kv) \right)\,,
\]
where $u,v$ are arbitrary vector fields and $ [. ,. ]$ denotes the commutator of two vector fields.

 On a  local coordinate chart $q = (q_1,\ldots,q_n) $ the  alternating (1, 2) Nijenhuis tensor  takes the
form
\[
(N_K)^i_{jk}=\sum_{\alpha=1}^n \left(
\dfrac{\partial K^i_k}{\partial q_\alpha}K^\alpha_j -
\dfrac{\partial K^i_j}{\partial q_\alpha}K^\alpha_k+
\left(\dfrac{\partial K^\alpha_j}{\partial q_k}-\dfrac{\partial K^\alpha_k}{\partial q_j}\right) K^i_\alpha
\right)
\]
The corresponding Haantjes tensor looks like
\[
 (H_K)^i_{jk}=\sum_{\alpha,\beta=1}^n  \left(K^i_\alpha K^\alpha_\beta (N_K)^\beta_{jk}  +
(N_K)^i_{\alpha \beta}K^\alpha_j K^\beta_k-
K^i_\alpha\left( (N_K)^\alpha_{\beta k} K^\beta_j+
 (N_K)^\alpha_{j \beta } K^\beta_k \right)\right)\,.
\]
Properties of these tensors are discussed in \cite{bog06,bog07}. In \cite{ts15,ts15a} we used components of these tensors for the description of the non-invariant Killing tensors
with $H_K\neq 0$.

\section{Three-dimensional Euclidean space}
\setcounter{equation}{0}
Let us consider Euclidean space $\mathbb R^3$ with Cartesian coordinates $q_1,q_2,q_3$ and metric
\bq\label{m-r3}
\mathrm g=\left(
            \begin{array}{ccc}
              1 & 0 &0 \\
              0 &1 & 0 \\
              0 & 0 & 1 \\
            \end{array}
          \right)\,.
\eq
The canonical basis of the  Killing vectors  consists of the translational Killing vectors
\[
X_1=\left(
    \begin{array}{c}
      1 \\
      0 \\
      0 \\
    \end{array}
  \right)\,,\qquad X_2=\left(
    \begin{array}{c}
      0 \\
      1 \\
      0 \\
    \end{array}
  \right)\,,\qquad X_3=\left(
    \begin{array}{c}
      0 \\
      0 \\
      1 \\
    \end{array}
  \right)
\]
and from the rotational Killing vectors
\[
R_1=\left(
    \begin{array}{c}
      0 \\
      -q_3 \\
      q_2 \\
    \end{array}
  \right)\,,\qquad R_2=\left(
    \begin{array}{c}
      q_3 \\
      0 \\
      -q_1 \\
    \end{array}
  \right)\,,\qquad R_3=\left(
    \begin{array}{c}
      -q_2 \\
      q_1 \\
      0 \\
    \end{array}
  \right)\,.
\]
Any Killing tensor of valence two  (\ref{k-tens})  in $\mathbb R^3$ is represented as a linear
combination of symmetric products of the basic Killing vectors
\bq\label{gen-k2}
K=\sum_{ij}^3 A^{ij}X_i\odot X_j + \sum_{ij}^3 B^{ij}X_i\odot R_j +  \sum_{ij}^3 C^{ij}R_i\odot R_j \,,
\eq
where matrices
\[
A=\left(
    \begin{array}{ccc}
      a_1 & \alpha_1 & \alpha_2 \\
      \alpha_1 &a_2 & \alpha_3 \\
      \alpha_2 & \alpha_3 & a_3 \\
    \end{array}
  \right)\,,\qquad
  B=\left(
                       \begin{array}{ccc}
                         b_{11} & b_{12} & b_{13} \\
                         b_{21} &b_{22} &b_{23} \\
                         b_{31} & b_{32} & b_{33} \\
                       \end{array}
                     \right)\,,\qquad
  C=\left(
      \begin{array}{ccc}
        c_1 & \gamma_1 & \gamma_2 \\
        \gamma_1 &c_2 & \gamma_3 \\
        \gamma_2 & \gamma_3 & c_3 \\
      \end{array}
    \right)
\]
depend on twenty-one parameters.  It is easy to observe  that entries of $K$ (\ref{gen-k2}) involve only the differences of the diagonal coefficients $b_{11}, b_{22}$ and $b_{33}$. Defining
\[\beta_1 = b_{22}- b_{33}\,,\qquad \beta_2= b_{33} - b_{11}\,,\qquad \beta_3 = b_{11} - b_{22}\,, \]
yields the constraint $\beta_1+\beta_2+\beta_3=0$, thereby showing that there are only twenty independent
parameters. It coincides with the Delong-Takeuchi-Thompson  formula
\bq\label{dtt}
d=\frac{1}{n}\binom{n+m}{m+1}\binom{n+m-1}{m}=\frac{1}{3}\binom{3+2}{2+1}\binom{3+2-1}{3}=20\,,
\eq
for the dimension of vector space of the Killing tensors of valency $m$ in $n$-dimensional Riemannian space, see \cite{smir05} for details.

In generic case$K$ (\ref{gen-k2}) is a completely non-invariant  tensor   with non-vanishing Haantjes torsion
\[
\mathcal L_{X_i}K\neq 0 ,\qquad \mathcal L_{R_i}K\neq 0\,,\qquad\mbox{and}\qquad   H_K\neq 0\,.
\]
Thus, we have two integrals of motion $H_{1,2}$ (\ref{ham-g}-\ref{int-st})
\bq\label{2-ham}
 H_1=\sum_{i,j=1}^3 \mathrm g^{ij}(q) p_i p_j+V(q)\,\qquad\mbox{and}\qquad H_2=\sum_{i,j=1}^3 K^{ij} p_i p_j+U(q)\,.
 \eq
 and
\begin{itemize}
  \item at $H_K=0$ we can construct third independent second-order polynomial in momenta $H_3$  using Eisenhart construction \cite{eis34};
  \item at $\mathcal L_{V}K=0$, where $V$ is a linear combination of translations $X_i$ and rotations $R_i$, we can construct a third independent linear integral of motion $H_3$ using Noether's theorem.
\end{itemize}
For instance, let us consider symmetric Killing tensor
 \[
 K=\left(
     \begin{matrix}
       a_1 + 2b_{13}q_2 + c_3q_2^2 & \alpha_3- b_{13}q_1 + b_{23}q_2 - c_3q_1q_2 & \alpha_2-\beta_1 q_2 \\ \\
       * & a_2- 2b_{23}q_1 + c_3q_1^2& \alpha_1 + \beta_1q_1 \\ \\
       * & * & a_3 \\
     \end{matrix}
   \right)
 \]
  which is invariant  under translation
 \[\mathcal L_{X_3} K=0.\]
 Because $H_{K_T}\neq 0$  we can not apply standard Eisenhart's construction. Nevertheless, using Noether's theorem we easily obtain desired integrals of motion in the involution
 \[
 H_1=\sum_{ij}^3\mathrm g^{ij}p_ip_j+V(q_1,q_2)\,,\qquad H_2=\sum_{ij}^3 K^{ij}p_ip_j+U(q_1,q_2)\,,\qquad H_3=p_3\,.
 \]
Similarly, we can construct integrable systems associated with rotational and helicoidal symmetries. So, integrable systems outside  Nijenhuis and Haantjes geometry exist.

At $H_K=0$, i.e. inside  Nijenhuis and Haantjes geometry,  two integrals of motion $H_{1,2}$  (\ref{2-ham})
\[
 H_1=\sum_{i,j=1}^3 \mathrm g^{ij}(q) p_i p_j+V(q)\,\qquad\mbox{and}\qquad H_2=\sum_{i,j=1}^3 K^{ij} p_i p_j+U(q)
 \]
completely define the integrable system with quadratic integrals of motion, see \cite{ben16} and more generic case in \cite{koz20}.

At $H_K\neq0$, i.e. outside  Nijenhuis and Haantjes geometry, we can also suppose that two integrals of motion completely determine an integrable system with "non-trivial " potentials.
Indeed, let us substitute generic Killing tensor $K$ (\ref{gen-k2}) depending on 20 parameters into the equation
\[\{H_1,H_2\}=0\qquad\Rightarrow\qquad d(KdV)=0\]
and try to solve the resulting equation imposing the following restrictions  on $K$
\[
\mathcal L_{X_i}K\neq 0 ,\qquad \mathcal L_{R_i}K\neq 0\,,\qquad\mbox{and}\qquad   H_K\neq 0\,.
\]
Here we do not use grading by momenta when we have to study the geodesic motion and only then add the suitable potential to each obtained geodesic.
We prefer to solve an equation on the potential $V$ depending on $20$ parameters which describe all the possible geodesics.

As a result, we obtain completely non-invariant Killing tensor
\bq\label{a-r3}
K=\left(
      \begin{array}{ccc}
        -q_2 &\dfrac{q_1}{2}  & 0 \\ \\
        \dfrac{q_1}{2} & 0 & - \dfrac{aq_3}{2}\\ \\
        0&  -\dfrac{aq_3}{2} &aq_2 \\
      \end{array}
    \right)\,,\quad a=1,-1/2,-2\,,
\eq
and potentials
\bq\label{int-pol}
 \mbox{first solution:}\qquad a=1,\qquad V(q)=\alpha\Bigl(q_1^4+6q_1^2q_3^2+q_3^4+12q_2^2(q_1^2+q_3^2)+16q_2^4\Bigr)\,,
  \eq
  and
 \bq
\label{int-d}
\mbox{second solution:}\qquad  a=-1/2,-2,\qquad\qquad  V(q)=\dfrac{\alpha (q_1^2+4 q_2^2+4 q_3^2)}{q_1^6}\,,\qquad\qquad
\eq
Other solutions are "trivial", i.e. potential is separable in Cartesian coordinates
\[V(q_1,q_2,q_3)=f_1(q_1)+f_2(q_2)+f_3(q_3)\]
and some of these separable potentials may be added to (\ref{int-pol}-\ref{int-d}), see \cite{ts15,ts15a}.

\subsection{First solution}
 If $a=1$, then solution of equation $d(KdV)$ is equal to
  \[
  V(q)=\alpha\Bigl(q_1^4+6q_1^2q_3^2+q_3^4+12q_2^2(q_1^2+q_3^2)+16q_2^4\Bigr)\,,
  \]
 solution of the corresponding equation $\{H_1,H_2\}=0$  has the form
  \[
U(q)=2\alpha q_2(q_1^2-q_3^2)(q_1^2+2 q_2^2+q_3^2)\,.
\]
The third integral of motion is the polynomial of the fourth-order in momenta
 \[ H_3=p_1^2p_3^2+2\alpha \sum_{i,j}^n S_{ij}(q) p_ip_j +4\alpha^2W(q)\]
 where
 \[
 S(q)=\left(
        \begin{array}{ccc}
          2q_2^2q_3^2 & -2q_1q_2q_3^2 & q_1q_3(q_1^2 + 4q_2^2 + q_3^2) \\ \\
          -2q_1q_2q_3^2 & 2q_1^2q_3^2 & -2q_2q_1^2q_3 \\ \\
          q_1q_3(q_1^2 + 4q_2^2 + q_3^2) & -2q_2q_1^2q_3 & 2q_1^2q_2^2 \\
        \end{array}
      \right)
  \]
 and
 \[
 W(q)=q_1^2q_3^2(q_1^2 + 2q_2^2 + q_3^2)^2\,.
 \]
 Thus, we have a non-trivial integrable system with two quadratic and one quartic invariants outside Nijenhuis and Haantjes geometry.

\subsection{Second solution}
 If $a=-1/2,-2$,  then Hamiltonian
  \[
  H_1=\sum_{ij=1}^3 \mathrm g^{ij}p_ip_j +V(q)=p_1^2+p_2^2+p_3^2+\dfrac{\alpha (q_1^2+4 q_2^2+4 q_3^2)}{q_1^6}\,,\]
 commutes with the following two  polynomials of second order in momenta
 \bq\label{ab-int}
 \begin{array}{rcl}
H_2&=&\displaystyle\sum_{i,j=1}^3 A^{ij}p_1p_j+U_2(q)=p_1J_{12}- 2p_3J_{23}-\dfrac{2\alpha q_2(q_1^2 + 2q_2^2 + 2q_3^2)}{q_1^6}\,,\\
\\
H_3&=&\displaystyle\sum_{i,j=1}^3 B^{ij}p_1p_j+U_3(q)=p_1J_{13} + 2p_2J_{23} - \dfrac{2\alpha q_3(q_1^2 + 2q_2^2 + 2q_3^2)}{q_1^6}\,,
\end{array}
\eq
 and with a component
\[
J_{23}=q_2p_3-q_3p_2
\]
of the angular momentum operator$J_{ij}=q_ip_j-q_jp_i$.

Here  $A$ and $B$ are the Killing tensor of valency two in $\mathbb R^3$ having non-vanishing Haantjes torsion
\[H_A\neq 0\qquad\mbox{and}\quad H_B\neq 0\,,\]
thus suggesting non-separability of the corresponding Hamilton-Jacobi equation in orthogonal curvilinear coordinate systems in $\mathbb R^3$.

Algebra of these integrals of the motion reads as
\[
\{H_1,H_2\}=\{H_1,H_3\}=\{H_1,J_{23}\}=0\,,\quad \{J_{23},H_2\}=H_3\,,\quad \{J_{23},H_3\}=-H_2
\]
and
\[
\{H_2,H_3\}=-4H_1\,J_{23}\,.
\]
Its second central element is the following polynomial of fourth order in momenta
\[
H_4=4H_1 J_{2, 3}^2 - H_2^2 - H_3^2\,.
\]
In this case, there are three sets of commuting integrals of motion
\[(H_1,H_2,H_4)\,,\qquad (H_1,H_3,H_4)\quad (H_1,J_{2,3},H_4)\]
which always involve two quadratic integrals and one quartic integral of motion.

Thus, we have a non-trivial superintegrable system with two quadratic and one quartic invariants outside Nijenhuis and Haantjes geometry.

\section{Four-dimensional Euclidean space}
\setcounter{equation}{0}
Let us consider Euclidean space $\mathbb R^4$ with Cartesian coordinates $q_1,q_2,q_3,q_4$ and metric
\[
\mathrm g=\left(
            \begin{array}{cccc}
              1 & 0 &0 &0\\
              0 &1 & 0 &0\\
              0 & 0 & 1 &0\\
              0&0&0&1\\
            \end{array}
          \right)
\]
According to  the Delong-Takeuchi-Thompson  formula (\ref{dtt}), the dimension of vector space of the Killing tensors of valency $4$ in $4$-dimensional Euclidean space is
\[
d=\frac{1}{n}\binom{n+m}{m+1}\binom{n+m-1}{m}=\frac{1}{4}\binom{4+2}{2+1}\binom{4+2-1}{3}=50\,.
\]
Unfortunately, we can not use modern computer algebra systems to calculate the generic solution of a system of partial differential equations in $\mathbb R^4$
\bq\label{4-meq}
d(KdV)=0\,,\qquad \mathcal L_{X_i}K\neq 0 ,\qquad \mathcal L_{R_i}K\neq 0\,,\qquad\mbox{and}\qquad   H_K\neq 0\,,
\eq
depending on  50 parameters.

So, let us study the following deformation of the Killing tensor (\ref{a-r3})
\[
K=\left(
      \begin{array}{cccc}
        -q_2 &\frac{q_1}{2}  & \scriptstyle 0 &\scriptstyle b_2 q_2+b_3 q_3+b_4 q_4+d_1\\ \\
        \frac{q_1}{2} & \scriptstyle 0 & - \frac{aq_3}{2}&\scriptstyle b_1 q_3-b_2 q_1+b_5 q_4+d_2\\ \\
        \scriptstyle 0&  -\frac{aq_3}{2} &aq_2 &\scriptstyle -b_1q_2 - b_3q_1 + b_6q_4+d_3\\ \\
        \scriptstyle b_2 q_2+b_3 q_3+b_4 q_4+d_1&\scriptstyle b_1 q_3-b_2 q_1+b_5 q_4+d_2&\scriptstyle -b_1q_2 - b_3q_1 + b_6q_4+d_3&
     \scriptstyle   - 2b_4q_1 - 2b_5q_2 - 2b_6q_3+d_4
      \end{array}
    \right)\,,
\]
 which is a partial solution of the Killing equation (\ref{k-tens}) depending on eleven parameters $a$, $b_1,\ldots,b_6$ and
 $d_1,\ldots d_4$. This deformation is invariant with respect to rotation
 \[
 \mathcal L_{R_{14}}K=0\,,\qquad R_{1,4}=q_1\frac{\partial}{\partial q_4}-q_4\frac{\partial}{\partial q_1}\,.
 \]

As a result,  we obtain three "nontrivial" solutions of the equations (\ref{4-meq}) associated with the Killing tensor
\bq\label{a-r4}
K=\left(
      \begin{array}{cccc}
        -q_2 &\dfrac{q_1}{2}  & 0 &0\\ \\
        \dfrac{q_1}{2} & 0 & - \dfrac{aq_3}{2}&bq_4\\ \\
        0&  -\dfrac{aq_3}{2} &aq_2 &0\\ \\
        0&bq^4&0&-2bq_2
      \end{array}
    \right)
\eq
at the special values of parameters
\begin{itemize}
  \item $a=1$ and $b=1/2$;
  \item $a=-1/2,-2$ and $b=1/2$;
  \item $a=-1/2,-2$ and $b=1$.
\end{itemize}
The corresponding integrable systems are discussed below.
\subsection{First solution}
 If  $a=1$ and $b=1/2$, then nonseparable in Cartesian coordinates solution of  (\ref{4-meq}) is
\bq\label{v4}
V(q)=\alpha \Bigl( q_1^4 + 12q_1^2q_2^2 + 6q_1^2q_3^2 + 2q_1^2q_4^2 + 16q_2^4 + 12q_2^2q_3^2 + 12q_2^2q_4^2 + q_3^4 + 6q_3^2q_4^2 + q_4^4 \Bigr)\,,
\eq
solution of the equation $\{H_1,H_2\}=0$ has the form
\bq\label{u4}
U(q)=2\alpha q_2(q_1^2 - q_3^2 + q_4^2)(q_1^2 + 2q_2^2 + q_3^2 + q_4^2)\,.\eq
Other integrals of motion read as
\[J_{14}=p_1q_4 - p_4q_1\]
and
\[
H_3=p_3^2(p_1^2 + p_4^2)+2\alpha \sum_{i,j=1}^4 S_{ij}(q) p_ip_j +\alpha ^2W(q)\,,
\]
where
\[
S(q)=\left(
    \begin{smallmatrix}
      2q_3^2(q_2^2 + q_4^2) &-2q_1q_2q_3^2  & q_1q_3(q_1^2 + 4q_2^2 + q_3^2 + q_4^2) & -2q_4q_1q_3^2 \\ \\
      -2q_1q_2q_3^2 & 2q_3^2(q_1^2 + q_4^2) & -2q_2q_3(q_1^2 + q_4^2) & -2q_2q_3^2q_4 \\ \\
      q_1q_3(q_1^2 + 4q_2^2 + q_3^2 + q_4^2) & -2q_2q_3(q_1^2 + q_4^2) &-2q_2q_3(q_1^2 + q_4^2) & q_3q_4(q_1^2 + 4q_2^2 + q_3^2 + q_4^2) \\ \\
      -2q_4q_1q_3^2 & -2q_2q_3^2q_4 & q_3q_4(q_1^2 + 4q_2^2 + q_3^2 + q_4^2) & 2q_3^2(q_1^2 + q_2^2) \\
    \end{smallmatrix}
  \right)
\]
and
\[
W(q)=4q_3^2(q_1^2 + q_4^2)(q_1^2 + 2q_2^2 + q_3^2 + q_4^2)^2\,.
\]
We can also add separable in Cartesian coordinates potentials to (\ref{v4}-\ref{u4})
\[
\begin{array}{rcl}
V_{ab}&=&\dfrac{c_1(q_1^2 + 4q_2^2 + q_3^2 + q_4^2)}{b} + c_2q_2 +\dfrac{c_3}{q_1^2} +\dfrac{c_4}{aq_3^2}+\dfrac{c_5}{bq_4^2}\,,\\
\\
U_{ab}&=& \dfrac{ c_1q_2(q_1^2-aq_3^2 - 2bq_4^2 )}{b}+ \dfrac{c_2(q_1^2-aq_3^2 + 2bq_4^2)}{4} -\dfrac{c_3q_2}{q_1^2}- \dfrac{c_4q_2}{q_3^2}  -\dfrac{2c_5q_2}{q_4^2}\,.
\end{array}
\]
This integrable system is a natural generalization of the 3D system (\ref{int-pol}).
\subsection{Second and third  solution}
If  $a=-1/2,-2$ and $b=1/2$, then Hamiltonian is equal to
\[  H_1=p_1^2+p_2^2+p_3^2+p_4^2 +\frac{\alpha (q_1^2 + 4q_2^2 + 4q_3^2 + q_4^2)}{(q_1^2 + q_4^2)^3}\,.\]
If  $a=-1/2,-2$ and $b=1$, then Hamiltonian reads as
\[  H_1=p_1^2+p_2^2+p_3^2+p_4^2 +
\frac{\alpha (q_1^2 + 4q_2^2 + 4q_3^2 + 4q_4^2)}{q_1^6}\,.
\]
 These Hamiltonian have quadratic invariants of the form (\ref{ab-int}), linear invariants which are components of the angular momentum tensor, and quartic invariant which is a central element of the corresponding algebra of invariants.

 For instance, at $a=-2$ and $b=1/2$ we have quartic invariant
 \[\begin{array}{rcl}
H_4&=&
(p_1^2 + p_4^2)\Bigl(J_{12}^2 + J_{13}^2 + J_{24}^2 + J_{34}^2\Bigr) - (p_2^2 + p_3^2)J_{14}^2\\
\\
&+&\displaystyle \frac{\alpha }{4(q_1^2 + q_4^2)^3}\sum_{i,j=1}^4 S_{ij}(q) p_ip_j +\alpha ^2W(q)\\
\end{array}
\]
where
\[
W(q)=\frac{\alpha (q_2^2 + q_3^2)(q_1^2 + 2q_2^2 + 2q_3^2 + q_4^2)^2}{4(q_1^2 + q_4^2)^6}
\]
and symmetric matrix $S$ is equal to
\[
S=\left(\begin{smallmatrix}
s_{11} &   -2q_1q_2(q_1^2 + 2q_2^2 + 2q_3^2 + q_4^2)  & -2q_1q_3(q_1^2 + 2q_2^2 + 2q_3^2 + q_4^2)&-q_1q_4(q_1^2 + 4q_2^2 + 4q_3^2 + q_4^2)\\
\\
 -2q_1q_2(q_1^2 + 2q_2^2 + 2q_3^2 + q_4^2)&4q_3^2(q_1^2 + q_4^2)&-4q_2q_3(q_1^2 + q_4^2)&-2q_2q_4(q_1^2 + 2q_2^2 + 2q_3^2 + q_4^2) \\
 \\
 -2q_1q_3(q_1^2 + 2q_2^2 + 2q_3^2 + q_4^2)&-4q_2q_3(q_1^2 + q_4^2)&4q_2^2(q_1^2 + q_4^2)&-2q_3q_4(q_1^2 + 2q_2^2 + 2q_3^2 + q_4^2)\\
 \\
 -q_1q_4(q_1^2 + 4q_2^2 + 4q_3^2 + q_4^2)&-2q_2q_4(q_1^2 + 2q_2^2 + 2q_3^2 + q_4^2)&-2q_3q_4(q_1^2 + 2q_2^2 + 2q_3^2 + q_4^2)&s_{44}\\
\end{smallmatrix}\right)\,,
\]
where
\[\begin{array}{rcl}
s_{11}&=&4q_1^2q_2^2 + 4q_1^2q_3^2 + q_1^2q_4^2 + 8q_2^4 + 16q_2^2q_3^2 + 8q_2^2q_4^2 + 8q_3^4 + 8q_3^2q_4^2 + q_4^4\,,\\
\\
s_{14}&=&q_1^4+8 q_1^2 q_2^2+8 q_1^2 q_3^2+q_1^2 q_4^2+8 q_2^4+16 q_2^2 q_3^2+4 q_2^2 q_4^2+8 q_3^4+4 q_3^2 q_4^2\,.
\end{array}
\]
At $a=-2$ and $b=1$ quartic invariant reads as
\[
H_4=p_1^2\Bigl(J_{12}^2 + J_{1 3}^2 + J_{14}^2\Bigr)+
\frac{2\alpha }{q_1^6}\sum_{i,j=1}^4 S_{ij}(q) p_ip_j +\alpha ^2W(q)\,,
\]
where
\[
W(q)=\frac{(q_2^2 + q_3^2 + q_4^2)(q_1^2 + 2q_2^2 + 2q_3^2 + 2q_4^2)^2}{4q_1^{12}}\,,
\]
and symmetric matrix $S$ is equal to
\[
S=\left(\begin{smallmatrix}
s_{11}& -q_2q_1(q_1^2 + 2q_2^2 + 2q_3^2 + 2q_4^2)&-q_3q_1(q_1^2 + 2q_2^2 + 2q_3^2 + 2q_4^2) &-q_4q_1(q_1^2 + 2q_2^2 + 2q_3^2 + 2q_4^2)\\
\\
-q_1q_2(q_1^2 + 2q_2^2 + 2q_3^2 + 2q_4^2)&2q_1^2(q_3^2 + q_4^2)&-2q_1^2q_2q_3& -2q_1^2q_2q_4\\
\\
-q_3q_1(q_1^2 + 2q_2^2 + 2q_3^2 + 2q_4^2) &-2q_1^2q_2q_3&2q_1^2(q_2^2 + q_4^2)&-2q_1^2q_3q_4\\
\\
-q_4q_1(q_1^2 + 2q_2^2 + 2q_3^2 + 2q_4^2)&-2q_1^2q_2q_4&-2q_1^2q_3q_4&2q_1^2(q_2^2 + q_3^2)\\
\end{smallmatrix}\right)\\
\]
where
\[
s_{11}=2(q_2^2 + q_3^2 + q_4^2)(q_1^2 + 2q_2^2 + 2q_3^2 + 2q_4^2)\,.
\]
These integrable systems are the generalization of the 3D superintegrable system (\ref{int-d}).

Using the result of the brute force solutions of the equations (\ref{4-meq}) we obtain a generalization of these superintegrable systems in $\mathbb R_3$ and $R^4$  to $\mathbb R^n$ which will be considered in the next Section.

\section{$n$-dimensional Euclidean space}
Below we change indexes in our previous formulae to get more compact expressions for additional integrals of motion.
\subsection{First family of superintegrable systems}
Let us consider Hamiltonian
\[
H_1=\sum_{i,j=1}^n\mathrm  g^{ij}p_ip_j+V(q_1,q_2,\rho)=\sum_{i=1}^n p_i^2+\alpha(4q_1^2+q_2^2+\rho) + 4\alpha(q_1^2 + q_2^2)\rho + 4\alpha q_1^2q_2^2\,,
\]
commuting with the second integral of motion
\bq\label{a-ten}
H_2=\sum_{i,j=1}^n K^{ij}p_ip_j+U(q_1,q_2,\rho)=\Bigl(\sum_{i=1}^n p_iJ_{i,1}\Bigr)-2p_2J_{2,1}+2\alpha q_1(\rho-q_2^2)\left(2q_1^2+q_2^2+\rho\right)\,.
\eq
Here
 \[\rho=q_3^2+\cdots+q_n^2\,,\]
 and  $J_{ij}$ are components of the angular momentum operator  $J$ in $T^*\mathbb R^n$
 \bq\label{ang-mom}
J=\left(
          \begin{array}{ccccc}
            0 &J_{1,2}  & J_{1,3} &\cdots & J_{1,n} \\
           J_{2,1} &0 & J_{2,3} &  &  \\
           J_{3,1} & J_{3,2} &0 & &\vdots \\
            \vdots &  &  & \ddots & J_{n-1,n} \\
           J_{n,1} &  &  \cdots & J_{n,n-1} & 0 \\
          \end{array}
        \right)\,,\qquad  J_{i,j}=q_ip_j-q_jp_i\,.
\eq
Integrals of motion $H_1$ and $H_2$ are in involution with the following quartic polynomial
\[
H_3=p_2^2\,\sum_{i=3}^n p_i^2+\sum_{i,j=1}^n S^{ij}(q_1,q_2,\rho)\,p_ip_j + W(q_1,q_2,\rho)\,,\quad
W=4\alpha^2 q_2^2\rho(\rho+2q_1^2+q_2^2)^2\,,
\]
where $S$ is a symmetric matrix with the following entries
\[\begin{array}{llll}
&S^{2,2}=4\alpha q_1^2\rho\,,\qquad  &S^{i,i}=4\alpha q_2^2(\rho+q_1^2 - q _ i^2)\,,\quad & i\neq 2;\\ \\
&S^{1,2}= -4\alpha q_1q_2\rho\,,\qquad &S^{i,2}=2\alpha q_1q_2(\rho+4q_1^2+q_2^2),\quad &i\neq 1,2\,,
\end{array}
\]
and in other cases
\[
S^{i, j} = -4\alpha q_ iq_ jq_2^2\,.
\]
Integrals of motion $H_1,H_2$ and $H_3$  commute with all the entries of the submatrix $\hat{J}$
obtained by deleting the first and second row and column in $J$ (\ref{ang-mom})
\[
\hat{J}=\left(
          \begin{array}{ccccc}
            0 &J_{3,4}  & J_{3,5} &\cdots & J_{3,n} \\
           J_{4,3} &0 & J_{4,5} &  &  \\
           J_{5,3} & J_{5,4} &0 & &\vdots \\
            \vdots &  &  & \ddots & J_{n-1,n} \\
           J_{n,3} &  &  \cdots & J_{n,n-1} & 0 \\
          \end{array}
        \right)\,,
\]
It means that $H_1,H_2$ and $H_3$  are invariants of  action of the group  $O(n-2)$ which is a subgroup of the corresponding  isometry group $\mathbb E(n)$.

Although the  $\hat{J}_{ij}$ are in involution with $H_1, H_2$ and $H_3$, they are not in involution with each other. As usual \cite{gram85}, we can introduce  $n-3$ integrals of motion in involution through
\[I_{k}=\sum_{j<k}^{n}\hat{J}^2_{j,k},\quad  k=3,\ldots,n\qquad\mbox{and}\qquad \lambda=\sum_{k=3}^n I_{k}\,.\]
It provides  integrability of at $n=3,4$ and superintegrability at $n>4$.

\subsection{Second family of superintegrable systems}
Let us consider Hamiltonian
\[
H_1=p_1^2+\cdots+p_n^2+\frac{\alpha}{r^2}\left(1+\frac{4\rho}{r}\right)\,,\quad r=q_1^2+\cdots+q_m^2\,,\quad \rho=q_{m+1}^2+\cdots+q_n^2\,,
\]
commuting with $n-1$ polynomials of second order in momenta
\[
H_k=\sum_{i=1}^m  p_iJ_{i k} + 2\sum_{j=m+1} p_j J_{jk} - \frac{2\alpha q_k}{r^2}\left(1+\frac{2\rho}{r}\right)\,,\qquad k=m+1..n\,,
\]
Here $m$ is arbitrary integer on the interval $0<m<n$.

Because Hamiltonian $H_1$ depends only on  with $r$ and $\rho$ it remains  invariant under action of the subgroups  $O(m)$ and $O(n-m-1)$ in the isometry group $\mathbb E(n)$.
As a result, $H_1$ commutes with entries of the truncated  angular momentum tensor
\[
\hat{J_{ij}}=\left(
              \begin{array}{cc}
                J_m & 0 \\
                0 & J_{n-m-1} \\
              \end{array}
            \right)
\]
The central element of the corresponding algebra of integrals of motion is  the following polynomial of fourth order in momenta
\[
C=4H_1\sum_{j>i}^{n-1}\hat{J}_{i, j}^2 -\sum_{k=2}^{n-1} H_k^2\,.
\]
which has to be included in all the sets of $n$ integrals of motion in the involution. Other central elements are the Hamiltonian and the Casimir element of $O(m)$.

For instance, at $n=5$ and $m=3$  Hamiltonian is equal to
 \[H_1=\sum_{i=1}^5 p_i^2 +\frac{\alpha}{(q_1^2 + q_2^2 + q_3^2)^2} \left(1+ \frac{4(q_4^2 + q_5^2)}{q_1^2 + q_2^2 + q_3^2}\right)\,.\]
It commutes with integrals of motion $H_4$ and $H_5$
\[
H_{k}=\sum_{i=1}^5  p_iJ_{ik} + 2\sum_{j=m+1}^5 p_j J_{j,k} - \frac{2\alpha q_k}{(q_1^2 + q_2^2 + q_3^2)^2}\left(1+\frac{2(q_4^2 + q_5^2)}{q_1^2 + q_2^2 + q_3^2}\right)\,,\qquad k=4,5\,,
\]
and with the following entries of the angular momentum vector  $J_{1,2}$, $J_{1,3}$ and $J_{2,3}$. Non-trivial elements of the  algebra of integrals of motion
are
\[
\{H_4,H_5\}=-4J_{4,5}H_1,\quad \{H_4,J_{45}\}=-H_5,\quad \{H_5,J_{45}\}=H_4\,,
\]
and
\[
\{J_{1,2},J_{1,3}\}=J_{2,3}\,,\qquad\{J_{1,3},J_{2,3}\}=J_{1,2}\,,\qquad \{J_{2,3},J_{1,2}\}=J_{1,3}\,.
\]
The central element of this algebra of integrals of motion has the form
\[
C=4H_1\left(J_{1,2}^2+J_{1,3}^2+J_{2,3}^2+J_{45}^2\right) - H_4^2 - H_5^2\,,
\]
which can be rewritten as
\[
\tilde{C}=4H_1J_{45}^2 - H_4^2 - H_5^2
\]
using other central elements $H_1$ and $J_{1,2}^2+J_{1,3}^2+J_{2,3}^2$.

\section{Conclusion}
Sometimes all the quadratic conservation laws for equations of mathematical physics can be determined using only two quadratic integrals of motion \cite{ben16, koz20}.
In this note, we proved that $n-1$ quadratic and one quartic constant of motion can be also determined using  only two quadratic integrals of motion
\[
 H_1=\sum_{i,j=1}^3 \mathrm g^{ij}(q) p_i p_j+V(q)\,\qquad\mbox{and}\qquad H_2=\sum_{i,j=1}^3 K^{ij} p_i p_j+U(q)\,,
 \]
where $K$ is the completely non-invariant Killing tensor outside  Nijenhuis and Haantjes geometry
\[H_K\neq 0\,.\]
In three-dimensional Euclidean space, these Killing tensors are characterized by the following condition on the off-diagonal  entries of Haantjes tensor $H_K$
\bq\label{3cond}
\left(H_K\right)^1_{23}=0\,,\qquad \left(H_K\right)^2_{31}\neq 0\,,\qquad \left(H_K\right)^3_{12}\neq 0\,,
\eq
up to permutation of indexes, see \cite{ts15,ts15a}. This experimental fact allows us to study similar Killing tensors in another three-dimensional space. For instance, we can prove that in the conformal Euclidean space with metric
\[
\hat{\mathrm g}=\dfrac{1}{1+\lambda f(q)} \,\left(
                                              \begin{array}{ccc}
                                                1 & 0 & 0 \\
                                                0 & 1 & 0 \\
                                                0 & 0 & 1 \\
                                              \end{array}
                                            \right)\,,\qquad \lambda\in \mathbb R,
\]
Killing tensors satisfying  (\ref{3cond}) do not exist at
\[ f(q)=q_1^2+q_2^2+q_3^2\]
and exist at
\[ f(q)=q_1^2+q_2^2+4q_3^2\,.\]
In the last case, we can construct at least two new integrable systems with 2 quadratic and 1 quartic invariants in the involution.

Similarly, we can study more complicated deformations of the flat metric. For instance, we suppose that satisfying equations  (\ref{3cond})
 Killing tensors
\begin{itemize}
  \item do not exist in the space 3-manifold for Schwarzschild coordinates
in the Schwarzschild spacetime when the metric is the following deformation of the flat metric $\mathrm g$
\[
\hat{\mathrm g}=\sigma\,\mathrm g\,,\qquad \sigma=\left(1+\frac{1}{2mr}\right)^4
\]
  \item exist in  the space 3-manifold for Boyer-Lindquist coordinates in the Kerr spacetime, because the corresponding metric is more complicated deformation of the flat metric
\[
\hat{\mathrm g}=\sigma\,\mathrm g+s\times s
\]
where $\sigma$ and $s$ are respectively a scalar and a one-form  depending on all the spherical coordinates $r$,  $\phi$ and $\theta$ \cite{coll02}.
\end{itemize}
Of course, it is more interesting to study non-invariant Killing or Yano-Killing tensors in  the four-dimensional  Schwarzschild spacetime and Kerr spacetime,  but it is a more complicated task.

In $4$-dimensional and $n$-dimensional Euclidean space we also have a few vanishing off-diagonal entries of Haantjes tensor $H_K$ whereas other off-diagonal entries do not equal to zero. Because the number of vanishing off-diagonal entries depends on a choice of local coordinates we need in the geometric description of the non-invariant Killing tensors with $H_K\neq 0$ suitable to construction of quartic invariants.

Similar to transformation from  the Nijenhuis conditions  (\ref{nts}) to more geometric Haantjes conditions (\ref{nh-zero} , we have to transform  (\ref{3cond}) to coordinate independent conditions working in any dimension. It is the main open theoretical problem appearing in our mathematical experiments with the non-invariant Killing tensors in Euclidean space.

The work was supported by the Russian Science Foundation (project 21-11-00141).

\end{document}